\documentstyle[aps,prd,psfig]{revtex}

\newcommand{\etal}{et al.~}
\newcommand{\gev}{{\rm GeV}}
\newcommand{\mev}{{\rm MeV}}
\newcommand{\kev}{{\rm KeV}}
\newcommand{\cm}{{\rm cm}}
\newcommand{\s}{{\rm s}}
\newcommand{\zC}{z_{\rm C}}
\newcommand{\zDC}{z_{\rm DC}}
\newcommand{\zeq}{z_{\rm eq}}
\newcommand{\zrec}{z_{\rm rec}}
\newcommand{\wb}{\omega_B}
\newcommand{\rg}{\rho_\gamma}
\newcommand{\rgo}{\rho_{\gamma 0}}
\newcommand{\nxo}{n_{X 0}}
\newcommand{\hef}{$^4$He~}

\begin{document}

\title{Cosmic Microwave Background constraint on residual 
annihilations of relic particles}

\author{Patrick McDonald,$^{1,2}$ Robert J. Scherrer,$^{1,3}$ 
and Terry P. Walker$^{1,3}$}

\address{$^1$Department of Astronomy, The Ohio State University,
Columbus, OH~~43210; mcdonald@astronomy.ohio-state.edu, 
scherrer,twalker@pacific.mps.ohio-state.edu}
\address{$^2$Department of Physics and Astronomy, University of Pennsylvania,
Philadelphia, PA~~19104}
\address{$^3$Department of Physics, The Ohio State University,
Columbus, OH~~43210}

\maketitle

\renewcommand{\baselinestretch}{1.3}

\begin{abstract}

Energy injected into the Cosmic Microwave Background 
at redshifts $z \lesssim 10^6$
will distort its spectrum permanently. 
In this paper we discuss the 
distortion
caused by annihilations of relic particles.  We use the observational
bounds on deviations from a Planck spectrum to constrain
a combination of annihilation cross section, mass, and 
abundance.  For particles with (s-wave) annihilation
cross section, 
$\left<\sigma \left| v\right|\right>\equiv \sigma_0$, the bound
is $f \left(m_X/\mev\right)^{-1}
\left[\left(\sigma_0/6\times 10^{-27} \cm^3~\s^{-1}\right)
\left(\Omega_{X \bar{X}}~h^2\right)^2 \right]< 0.2$, where 
$m_X$ is the particle mass, $\Omega_{X \bar{X}}$ is the fraction of 
the critical density the particle and its antiparticle 
contribute if they survive to the present time, 
$h=H_0/100~{\rm km~s^{-1} Mpc^{-1}}$,
$H_0$ is the Hubble constant, and $f$ is the fraction of the
annihilation energy that interacts electromagnetically.  
We also compute the less stringent limits for p-wave annihilation. 
We update other bounds on residual annihilations and compare 
them to our CMB bound.

\end{abstract}


\twocolumn

\section{INTRODUCTION}

Newly proposed particles, especially dark matter candidates, must
evade an ever growing array of empirical constraints, ranging
from bounds obtained in terrestrial laboratory experiments 
(e.g., reference \cite{groom}), 
to limits on new cooling sources in stars in our galaxy \cite{raffelt}, 
to constraints on the expansion rate of the universe during 
Big Bang Nucleosynthesis (BBN), when the universe
was only a few minutes old \cite{olive}.  In this paper we add
another constraint to the list, based on the effect of 
annihilations of relic particles on the cosmic microwave background
(CMB).  

The CMB energy spectrum provides 
a direct probe of the early Universe at redshifts as high as 
$z\sim 2 \times 10^6$.  Below this redshift, distortions of 
the spectrum generally cannot be thermalized and are 
observable today.  The FIRAS
(Far Infrared Absolute Spectrophotometer) instrument on the 
COBE (Cosmic Background
Explorer) satellite measured the spectrum and found it to have 
a Planck distribution 
to within a few hundredths of a percent \cite{fixsen}.
This measurement places strong upper bounds on any energy injection 
into the CMB after the thermalization redshift 
(e.g., references \cite{burigana,hua}).    

Decays of relic particles have been considered as a source of CMB 
distortions \cite{ellis},
but particle annihilations have not.  Annihilations are typically
ignored because the classic WIMP (weakly interacting massive particle)
dark matter candidates have mass $m_X\gtrsim \gev$ and their 
annihilations ``freeze out''
at $T_F \sim m_X/20 \gtrsim 50~\mev$
\cite{lee}, long
before the time when the CMB becomes vulnerable to distortion
(at $T\simeq 0.5~\kev$).
Freezeout is defined as the time when annihilations cease to change
significantly the number density of the particle; however,
annihilations continue eternally at some small rate and their products
can distort the CMB spectrum if they interact 
electromagnetically.  

 The effects of residual annihilations have been considered in 
a few other contexts:
Reno and Seckel 
\cite{reno}, Hagelin, Parker, and Hankey \cite{hagelin}, and
Frieman, Kolb, and Turner \cite{frieman} computed the effect of 
annihilation products on the primordial element abundances.
Cline and Gao \cite{cline} and Gao, Stecker, and Cline \cite{gao} 
considered
the possibility of observing directly the photons from annihilations
at cosmological distances.  Bergstrom and Snellman \cite{bergstrom}, 
Rudaz \cite{rudaz},
and Rudaz and Stecker 
\cite{rudsteck} discuss the detectability of a line source
from annihilations to photons within the Milky Way halo.
Jungman, Kamionkowski, and Griest
\cite{jungman} conclude that
halo annihilations into particles other than photons probably cannot
be used to place general constraints on particle candidates because of
astrophysical uncertainties.

In this paper we compute the energy injected into the CMB by annihilating
particles as a function of their mass and annihilation rate 
(i.e., the product
of cross section and abundance squared).  We derive constraints on
the particle properties by comparison with the observed limits on 
chemical potential
($\mu$) distortions, and Compton-$y$ distortions (\S II).  
We compare these constraints
to similar constraints obtained from the production of deuterium
by photodissociation of 
primordial helium (\S III.A), and from the diffuse photon 
background produced after recombination by extragalactic annihilations 
(\S III.B), and annihilations in the Milky Way halo (\S III.C).

\section{DISTORTIONS OF THE CMB ENERGY SPECTRUM}

We consider first the effect of annihilation products on the CMB energy
spectrum.  The distortion of the spectrum takes place in two steps:  
first the high energy annihilation products rapidly dissipate their 
energy into the 
background photons and electrons, and then the low energy background
evolves more slowly in an effort to restore the Planck spectrum.
 The permanence of distortions produced after $z\simeq 10^6$ is
simple to understand in the following way:  
A Planck spectrum with a given photon number density
must have a specific energy density.  For $z\lesssim 10^6$, photon 
nonconserving processes (double Compton scattering and 
bremsstrahlung) are inefficient in the background plasma.  
Therefore, if energy
is injected into the CMB but not the the correct number of photons, a Planck
spectrum cannot be restored.
We now discuss in more detail the form of the 
distortions produced in different
redshift intervals.

Down to recombination at $\zrec \simeq 1100$, photons 
with $E_\gamma\gtrsim 5~\kev$
quickly cool and produce heated electrons  
by Compton scattering ($\gamma e\rightarrow \gamma e$) or pair production 
on ions ($\gamma N \rightarrow N e^+ e^-$)
(all photons with $E_\gamma > 1~\kev$
can cool if $z\gtrsim 3500$).  The heated electrons 
quickly dissipate their energy by inverse Compton 
scattering on the huge number of CMB photons. 
This process produces a distorted spectrum with phase-space 
distribution
\begin{equation}
f(x,y)\simeq f(x,0) + y \frac{x e^x}{\left(e^x -1\right)^2}
\left[ \frac{x}{\tanh(x/2)}-4\right]~,
\label{compspec}
\end{equation}
where $f(x,0)=1/(e^x-1)$ is the Planck distribution, $x=E/T$, 
and the Compton-$y$ parameter is assumed to satisfy $y<<1$  
\cite{bernstein}.  In our case, where there
is ample time for all of the input energy to be transferred to the
CMB, the relation between $y$ and the input energy can be found
by integrating Equation (\ref{compspec}) to find the energy density
as a function of $y$.  The result is
$4~ y=\delta \rg/\rg$, where $\delta \rg$ is
the injected energy and $\rg$ is the energy of the CMB photons  
(see reference \cite{sunyaev} for a general review of CMB 
distortions or reference \cite{bernstein} for a more detailed 
discussion of the Compton-$y$ distortion).  
Analysis of the COBE FIRAS data set gives 
$\left|y\right| < 1.5 \times 10^{-5}$ \cite{fixsen}.   

The Compton-$y$ distortion will be preserved to the present time if
it is produced after $\zC \simeq 5.4\times 10^4 \wb^{-1/2}$
(where $\wb\equiv \Omega_B h^2/0.02$),
the redshift above which
elastic Compton scattering would effectively
redistribute energy between CMB photons,
converting a Compton-$y$ distorted spectrum into a Bose-Einstein 
spectrum with distribution 
\begin{equation}
f(x,\mu) = \frac{1}{\exp(x+\mu)-1}~,
\label{bespec}
\end{equation}
where $\mu$ is the chemical 
potential.
Assuming the input number of photons is negligible, which will 
always be true in this paper, the chemical potential is
 $\mu=1.4~\delta \rg / \rg$ \cite{hub}.
The FIRAS limit on this type of distortion is 
$\left|\mu\right|<9\times 10^{-5}$
\cite{fixsen}. 

Equation (\ref{bespec}) describes the equilibrium
distribution for a fixed total number of photons and amount of 
energy.  
For $z\gtrsim \zDC \simeq 2.1 \times 10^6 
\wb^{-2/5}$, photons are produced 
effectively by double Compton 
scattering ($e\gamma\rightarrow e\gamma\gamma$) so a Planck 
spectrum ($\mu=0$) can be restored for arbitrary energy input
\cite{hub}. 
Production of photons by bremsstrahlung is already ineffective at 
$z \sim \zDC$
(for the observed baryon density).

To summarize:  annihilations occurring at $z\gtrsim 2.1\times 10^6$
will be unobservable, annihilations in the range
$5.4\times 10^4 \lesssim z \lesssim 2.1\times 10^6$ will produce a 
Bose-Einstein spectrum with chemical potential $\mu$,
and annihilations in the range $1100 \lesssim z \lesssim 5.4\times 10^4$ 
will produce a Compton-$y$ distortion.  Annihilations at $z<\zrec$ 
will not significantly affect the CMB energy spectrum, but can be
observable in the diffuse photon background.

We now compute the fractional energy injection $\delta \rg / \rg$ (where
$\rho_\gamma$ is the energy density of the CMB) from annihilations of
particle species $X$.
We will assume throughout this paper that 
particle $X$ and antiparticle $\bar{X}$ are not identical (we
discuss below how our final constraints are strengthened 
in the case where the particle is its own antiparticle). 
We also assume that $n_X \equiv n_{\bar{X}}$, where
$n_X$ ($n_{\bar{X}}$) is the number density of $X$ ($\bar{X}$) 
(if there is a
significant asymmetry the relic density of the less numerous 
particle will usually be negligibly small 
\cite{scherrer,escherrer}).
The energy produced per annihilating
particle is
\begin{equation}
E_a \equiv f~m_X~,
\end{equation}
where $m_X$ is the mass of the particle
and $f\leq 1$ is the mean fraction of the annihilation energy that
interacts electromagnetically, and 
the rate of annihilations per unit volume is
\begin{equation}
\Gamma_a = \left< \sigma_a \left| v \right| \right> n_X n_{\bar{X}}~,
\end{equation}
where $\left< \sigma_a \left| v \right| \right>$
is the cross section.
A useful parameterization for the cross section is
$\left< \sigma_a \left| v \right| \right> \equiv 
\sigma_0~(T/m_X)^n$, where $n=0$ for s-wave annihilation 
(e.g., Dirac neutrinos) and
$n=1$ for p-wave annihilation (e.g., Majorana particles annihilating
into much lighter fermions, see reference \cite{goldberg}). 
We integrate the energy injection rate from time $t_1$ to $t_2$
as follows:
\begin{eqnarray}
\frac{\delta\rg}{\rg} &=&
\int_{t_1}^{t_2} \frac{\dot{\rho}_{\rm ann}}{\rg} dt =
\int_{t_1}^{t_2} \frac{2 E_a \Gamma_a}{\rgo~(1+z)^4} dt \nonumber \\ &=&
\frac{2 E_a}{\rgo}\int_{t_1}^{t_2} 
\sigma_0 \left(\frac{T}{m_X}\right)^n \nxo^2 (1+z)^2 dt \nonumber \\
 &=&
 \frac{4~t_\star~
E_a \sigma_0~\nxo^2}{\rgo} \left(\frac{T_0}{m_X}\right)^n
\int_{z_2}^{z_1} \left(1+z\right)^{n-1} dz ~,
\label{erateq}
\end{eqnarray}
where $\dot{\rho}_{\rm ann}=2 ~E_a \Gamma_a$ is the rate of energy 
injection by
annihilations, $\rgo$ is the present energy density in the CMB,
$T_0$ is the present CMB temperature,
$\nxo\equiv n_X\left(z\right)~(1+z)^{-3}$, and 
$t_\star \equiv 2.4\times 10^{19}$s. 
We have assumed the universe is radiation dominated 
(this assumption 
will be accurate enough at the redshifts relevant to our 
calculation) so that
$t\simeq 0.301~g_\star^{-1/2} m_{Pl}~T^{-2} \equiv t_\star (1+z)^{-2}$, 
where
$m_{Pl}$ is the Planck mass, $g_\star=3.36$ is the number of 
effectively massless degrees of freedom at $T<<\mev$, and $t_\star$
is fixed by the CMB temperature measured at $z=0$. 
The result for $n=0$ is
\begin{equation}
\frac{\delta \rg}{\rg}=A ~\ln\left(\frac{z_1}{z_2}\right) ~,
\end{equation}
and for $n=1$
\begin{equation}
\frac{\delta\rg}{\rg}=
A ~\frac{T_1-T_2}{m_X}~,
\end{equation}
where
\begin{equation}
A\equiv \frac{4~t_\star~E_a ~\sigma_0~\nxo^2}{\rgo} ~,
\end{equation}
and $T_i\equiv T(z_i)$. 

We use $z_1=\zDC$ and $z_2=\zC$ to find the chemical potential
$\mu=1.4~\delta \rg/ \rg$ (this assumes a negligible increase in the
number of photons, which is valid even in the case of an electromagnetic
cascade), with the
results:
\begin{eqnarray}
\mu &=& 5.1~A = 2.9\times 10^{-4}f \left(\frac{m_X}{\mev}\right)^{-1}
\nonumber \\ & & \times
\left[\left(\frac{\sigma_0}{6\times 10^{-27} \cm^3 \s^{-1} }\right)
\left(\Omega_{X\bar{X}}~h^2\right)^2 \right]~~
({\rm for}~ n=0),
\end{eqnarray}
and
\begin{eqnarray}
\mu &=& 2.0\times 10^6~A \frac{T_0}{m_X}= 
2.7\times 10^{-8}  f \left(\frac{m_X}{\mev}\right)^{-2}
\nonumber \\ & & \times
\left[\left(\frac{\sigma_0}{6 \times 10^{-27}\cm^3\s^{-1}}\right)
\left(\Omega_{X\bar{X}}~h^2\right)^2 \right]~~({\rm for}~ n=1),
\end{eqnarray}
where $\Omega_{X\bar{X}} h^2 = 94  \left(m_X/\mev\right) 
\left(n_{X 0}+n_{\bar{X} 0}\right)/\cm^{-3}$.  

Observationally $\left|\mu\right|<9\times 10^{-5}$
\cite{fixsen} so, for $n=0$, we
have the bound
\begin{eqnarray}
f \left(\frac{m_X}{\mev}\right)^{-1}
\left[\left(\frac{\sigma_0}{6\times 10^{-27} \cm^3 \s^{-1} }\right)
\left(\Omega_{X \bar{X}}~h^2\right)^2 \right]
\nonumber \\ < 0.3~.
\label{n0mubound}
\end{eqnarray}
Similarly, for $n=1$ we find
\begin{eqnarray}
f \left(\frac{m_X}{\mev}\right)^{-2}
\left[\left(\frac{\sigma_0}{6 \times 10^{-27}\cm^3\s^{-1}}\right)
\left(\Omega_{X \bar{X}}~h^2\right)^2 \right] \nonumber \\
<3.3\times 10^3~.
\label{n1mubound}
\end{eqnarray}

We have chosen to scale the cross section as 
$(\sigma_0/6\times 10^{-27} \cm^3 \s^{-1})$ 
because this is approximately the value at which a non-relativistic
particle will have relic density $\Omega_{X \bar{X}} h^2 = 1$ 
if it freezes out 
as a result of its annihilations while in thermal equilibrium.  
The reader should keep in mind that $\Omega_X$ is not
necessarily the contribution  of $X$
to the present energy density if, for example, the particle decays
subsequent to distorting the CMB. 

Note that the constraint for $n=1$ is roughly $m_X/T_{\rm DC}$
times weaker than the constraint for $n=0$,
where $T_{\rm DC}$ is the temperature at $\zDC$. 
Constraints on $n=1$ particles that fall out of kinetic equilibrium
at $T_K>T_{\rm DC}$ can be roughly estimated from the given 
$n=1$ constraints by 
multiplying by a factor $T_K/T_{\rm DC}$ (see reference 
\cite{hannestad}).

The Compton-$y$ distortion can be obtained by changing the
limits of integration in Equation (\ref{erateq}) from $\zDC$ and
$\zC$ to $\zC$ and $\zrec$.
For simplicity, we replace the lower limit $\zrec$ by  
the redshift of matter-radiation equality,  
$\zeq \simeq 2.5\times 10^4 ~\Omega_0 h^2$ (where $\Omega_0$ is the 
present density in matter), and we ignore deviations from 
$t\propto T^{-2}$.  The precise value of this lower limit is
not important (because of the log dependence) 
so we assume $\zeq = 3200$.

For $n=0$, 
\begin{equation}
y\simeq \frac{\mu}{1.4}\frac{1}{4}\frac{\ln\left(\zC/\zeq\right)}
{\ln\left(\zDC/\zC\right)}=0.15\mu~,
\end{equation}
while the observational bound is $\left|y\right|<1.5\times 10^{-5}$
\cite{fixsen},
giving a bound essentially identical to Equation (\ref{n0mubound}).
For $n=1$, $y\simeq 0.005~\mu$, so the bound will be significantly
weaker than Equation (\ref{n1mubound}).

In the case of $n=0$, we predict that the $\mu$ and $y$ distortions
will always appear together, with the relationship $y\simeq 0.15~\mu$.
Fixsen \etal \cite{fixsen} do not give joint constraints on $\mu$
and $y$ so we perform our own linear fit to the data in their Table 4,
using the formula
\begin{eqnarray}
I_0(\nu)-B_{\nu}(T_0)-G_0 g(\nu)=
\Delta T~\frac{\partial B_{\nu}}{\partial T}+
G_1 g(\nu)+ \nonumber \\
1.4~\frac{\delta \rho}{\rho} \left(
\frac{\partial S_c}{\partial \mu}+0.15 \frac{\partial S_c}{\partial y}
\right)~,
\end{eqnarray}
where $I_0(\nu)-B_{\nu}(T_0)-G_0 g(\nu)$ is taken from 
reference \cite{fixsen} 
(along with the necessary error bars), 
$g(\nu)= \nu^2 B_{\nu}(9~K)$ is the galactic 
contamination model used by \cite{fixsen}, and 
$\partial S_c / \partial p$ is the deviation
from a blackbody as parameter $p$ is varied.  We constrain jointly
the parameters $\Delta T$, $G_1$, and $\delta \rho / \rho$.  Based 
only on statistical errors, the
95\% confidence upper bound on the energy injection is 
$\delta \rho / \rho < 1.6\times 10^{-5}$.  
Our analysis is complicated by the significant systematic errors, 
$1\times 10^{-5}$ and $4\times 10^{-6}$, quoted by \cite{fixsen} 
for $\mu$ and $y$, respectively.  It is not clear how these should
be combined with the statistical error, so we choose arbitrarily to
add the two systematic errors linearly, and then add the result in
quadrature to the statistical error, to find a final 95\% confidence 
upper bound $\delta \rho / \rho < 4.1\times 10^{-5}$.
We finally obtain the bound
\begin{eqnarray}
f \left(\frac{m_X}{\mev}\right)^{-1}
\left[\left(\frac{\sigma_0}{6\times 10^{-27} \cm^3 \s^{-1} }\right)
\left(\Omega_{X \bar{X}}~h^2\right)^2 \right]
\nonumber \\ < 0.2~.
\label{ncombmubound}
\end{eqnarray}
The combined constraints are plotted as the solid lines in 
Fig. \ref{config} (for $n=1$ we just use the $\mu$ constraint).

We truncated Fig. \ref{config} at $m_X=1~\gev$, where hadronic
interactions become important; however, the CMB
bound should continue to apply at higher energies, as long as $f$ 
is computed to account for non-interacting annihilation products.  
Even neutral annihilation products
can contribute to the CMB distortion if
they are coupled to the plasma in any way, for example, 
if they heat the background protons through
hadronic interactions (see references \cite{reno,hagelin}
for discussions of hadronic interactions during this epoch).
This is not generally true for the other constraints that we
review in \S III.

These constraints apply to a particle that is distinct from its 
antiparticle; however, to convert to the case where particle and 
antiparticle are equivalent, it is only necessary to make the
substitutions $\Omega_{X \bar{X}} \rightarrow \Omega_X$, and 
$\sigma_0/6\times 10^{-27}{\rm cm^3 s^{-1}}\rightarrow  \sigma_0/
3\times 10^{-27}{\rm cm^3 s^{-1}}$.  The change in cross section 
normalization cancels the increase in the annihilation rate at fixed
total contribution to the critical density
(i.e., for the inequivalent case
$\Gamma_a\propto n_X n_{\bar{X}} \propto
\Omega_{X \bar{X}}^2/4$, but for the equivalent case
$\Gamma_a \propto n_X^2/2 \propto \Omega_X^2/2$).
The scale $3 \times 10^{-27} {\rm cm^3 s^{-1}}$ for the cross section is 
natural also because it gives $\Omega_X h^2 \simeq 1$.
The same substitutions can be used to convert any of the constraints
in \S III.
  
\section{COMPARISON WITH OTHER CONSTRAINTS ON RESIDUAL ANNIHILATIONS}
In this section we compute bounds, in a form similar to 
Equation (\ref{n0mubound}), from the production of deuterium by
photodissociation of \hef (\S III.A), from the diffuse photon 
background produced by extragalactic annihilations at $z<\zrec$ (\S III.B),
and from the diffuse photon background produced by annihilations in 
our galaxy (\S III.C).  
We restrict our attention to $n=0$ because the $n=1$ bound is 
very weak in all cases. 

\subsection{Photodissociation of $^4$He}

At roughly the same time that the CMB energy spectrum becomes 
vulnerable to distortion by energy injection, primordially 
produced \hef nuclei become vulnerable to photodissociation
by high energy annihilation products. 
At earlier times
the nuclei were protected from destruction because photons with
high enough energy ($E_\gamma \gtrsim 20~\mev$) to destroy nuclei
instead
pair produce or elastic scatter on the CMB photons 
($\gamma \gamma\rightarrow e^+ e^-$, or $\gamma \gamma\rightarrow
\gamma \gamma$). 
This shielding is effective for $E_\gamma
\gtrsim m_e^2/44~T$, where the numerical factor accounts for the
fact that photons in the high energy tail of the CMB spectrum
are still very numerous (see reference \cite{sarkar}, 
and references therein, for a full discussion).  
Once $m_e^2/44~T \gtrsim 20~\mev$, 
annihilation products can dissociate \hef, either directly, or 
indirectly through a cascade when $m_X \gtrsim m_e^2/44~T$.

Bounds on energy injection can be derived by considering the 
$^3$He or the D produced when \hef nuclei are dissociated, and 
requiring that the amount created is not greater than the observationally
inferred primordial abundances.
(bounds from direct photodissociation of deuterium extend
to slightly lower annihilation energy but are significantly weaker
\cite{ellis}).
Previous analyses \cite{frieman,kolb} 
used $^3$He+D because
the upper limit on the primordial abundance of D was poorly
known; 
however, there is considerable uncertainty in the post-big bang 
production and destruction of $^3$He so we will use D because 
its abundance is now more
robustly measured in QSO absorption systems 
\cite{olive}, with a very 
conservative upper bound $n_D/n_H \equiv D/H<3\times 10^{-4}$ 
(high D) \cite{webb}, or, more probably,
$D/H <4\times 10^{-5}$ (low D) \cite{burles}. 

Protheroe, Stanev, and Berezinsky \cite{protheroe}
presented a detailed computation of the (redshift dependent) 
quantity of D produced for a given amount of injected
energy, $N_D(z)$.  With this input, the ratio $D/H$ is given 
by
\begin{equation}
\frac{D}{H}=\int \frac{N_D(z)~2 E_a \left< \sigma_a \left| v \right| \right>
n_X^2}{n_H}dt ~.
\end{equation}
For $n=0$ this is
\begin{eqnarray}
\frac{D}{H} &=& 2 E_a \int \frac{N_D(z) \sigma_0 \nxo^2 (1+z)^6}
{n_{H 0} (1+z)^3}\frac{2~t_\star}{(1+z)^3}dz \nonumber \\
&=&
\frac{4~t_\star  E_a \sigma_0 \nxo^2}{n_{H 0}}
\int N_D(z) dz = \frac{9800~t_\star E_a \sigma_0 \nxo^2}{n_{H 0}}~,
\end{eqnarray}
where we have used the results in \cite{protheroe} and evaluated the
integral numerically.
The constraint on annihilations of particles with $m_X\gtrsim 26~\mev$
(the energy needed to dissociate \hef into D) is
\begin{eqnarray}
f~\omega_B^{-1} \left(\frac{m_X}{\mev}\right)^{-1}
\left[\left(\frac{\sigma_0}{6\times 10^{-27} \cm^3 \s^{-1} }\right)
\left(\Omega_{X \bar{X}}~h^2\right)^2 \right]
\nonumber \\ < 1.3~({\rm high~D}),~{\rm or}
\nonumber \\
< 0.2~({\rm low~D})~,
\label{n0Dbound}
\end{eqnarray}
weaker than the CMB bound if the high deuterium abundance is used, but
similar for the low deuterium abundance
(recall $\wb = \Omega_B h^2/0.02 \simeq 1$). 
These constraints are plotted as the long-dashed lines in Fig. \ref{config}.

This calculation is not exactly correct unless $m_X >> 26~\mev$ because
our use of the results in \cite{protheroe} assumes that the annihilation 
products are energetic enough to produce a cascade.  A more careful 
calculation would weaken the bound slightly for $m_X\sim 26~\mev$
\cite{frieman}.

 Our constraint appears to be several orders of magnitude stronger
than the constraint described by Equation (9) in reference 
\cite{kolb}.
They used the observational bound $(^3 He+D)/H\lesssim 1.1\times 10^{-4}$,
which, considering that $^3$He is produced $\sim 24$ times more 
efficiently than D \cite{protheroe}, 
should give a bound somewhat stronger than our 
Equation (\ref{n0Dbound}).
We believe that the discrepancy is the result of a numerical error in 
their calculation.  Our result for the $(^3 He+D)/H$ calculation 
agrees with \cite{frieman}; however, we do not use this result
for our annihilation constraint because the deuterium constraint should
be more reliable \cite{olive}. 

\subsection{Extragalactic $\gamma$ Background}

Recently, Kribs and Rothstein \cite{kribs} used the observed $\gamma$-ray 
background to constrain late decaying particles.
Gao, Stecker, and Cline \cite{gao} computed the expected $\gamma$-ray
flux from annihilations of a possible lightest supersymmetric
particle. 
Here we generalize this annihilation calculation to match the form
of the constraint derived from the CMB.
Unlike the previous constraints,
the constraint we compute now only applies to annihilations 
directly to two photons,
which typically will be sub-dominant for particles with mass greater
than the electron mass (assuming particle $X$ does not couple directly
to photons).  Therefore, we will only calculate the constraint
for $m_X< 0.5~\mev$ (constraints for general annihilation products 
would be strongly model dependent). 

Temporarily ignoring the possibility that photons produced
after recombination (at $\zrec \simeq 1100$) are
absorbed on their way to the observer,
the observed energy flux at present from extragalactic
annihilations to photons is 
\begin{eqnarray}
\frac{dJ}{dE d\Omega}&=&
\frac{2~c}{4 \pi} \int_0^{z_{rec}}
dt~ f_\gamma m_X ~ \Gamma_a(z) \nonumber \\ & & \times
\delta
\left[E_o \left(1+z\right)-m_X\right]~(1+z)^{-3}~,
\end{eqnarray}
where $f_\gamma$ is the fraction of annihilations that produce
a pair of photons with $E_\gamma=m_X$.
Using $dt/da=3~t_0 a^{1/2}/2=H_0^{-1} a^{1/2}$ for an 
Einstein-de Sitter 
universe, and $\delta\left(E_o a^{-1}-m_X\right)=
a^2 \delta\left(a-E_o/m_X\right)E_o^{-1}$, we find
\begin{eqnarray}
\frac{dJ}{dE d\Omega} &=& \frac{2~c}{4 \pi H_0} f_\gamma
\sigma_0~\nxo^2 \left(\frac{m_X}{E_o}\right)^{3/2} \nonumber \\
&\simeq& \frac{3.9\times 10^{-4}}{{\rm cm^2~ s ~sr}} f_\gamma
\left(\frac{m_X}{\mev}\right)^{-2}\left(\frac{m_X}{E_o}\right)^{3/2} 
\nonumber \\ & & \times 
\left[\left(\frac{\sigma_0}{6\times 10^{-27} \cm^3 \s^{-1}}\right)
\left(\Omega_{X \bar{X}} h^2 \right)^2\right]~,
\end{eqnarray}
where we have taken $n=0$.
The photons we are considering,
with energy $E_\gamma<0.5~\mev$, will in fact lose most
of their energy by Compton scattering
if they are produced at $z\gtrsim 200$ (see reference \cite{zdziarski}),
so the maximum redshift factor is $m_X/E_o=200$.

We derive a bound on annihilations by requiring that the predicted photon
background is not greater than the observed one.
After considering observations of the photon background at all relevant
energies (see references \cite{skreekumar,chen,kt}), we find that the
ASCA data in the energy 
range $0.8~\kev \lesssim E_o \lesssim 2.5~\kev$
provides the best constraint on the annihilating particles
that we consider.  
The background in this range is conservatively
bounded by $dJ/dE d\Omega<0.36
\left(E_o/\mev\right)^{-0.58}{\rm cm^{-2} s^{-1} sr^{-1}}$ 
\cite{chen}.

For $160~\kev \lesssim m_X < 500~\kev$, the best constraint is derived from
photons produced at $z\simeq 200$.
Setting $E_o=m_X/200$, we obtain the bound   
\begin{eqnarray}
f_\gamma \left(\frac{m_X}{\mev}\right)^{-1.42}
\left[\left(\frac{\sigma_0}{6\times 10^{-27} \cm^3 \s^{-1}}\right)
\left(\Omega_{X \bar{X}} h^2 \right)^2\right] \nonumber \\
\lesssim 7.6~.
\end{eqnarray}
The best bound for lower annihilation
energies comes from observations at $E_o\simeq 0.8~\kev$:
\begin{eqnarray}
f_\gamma \left(\frac{m_X}{\mev}\right)^{-1/2}
\left[\left(\frac{\sigma_0}{6\times 10^{-27} \cm^3 \s^{-1}}\right)
\left(\Omega_{X \bar{X}} h^2 \right)^2\right] \nonumber \\
\lesssim 1.3~.
\end{eqnarray}
Figure \ref{config} shows these constraints as the dashed line.

\subsection{Annihilations in the Milky Way Halo}

The observability of annihilations to photons in our own galaxy has 
been discussed in many papers, including \cite{bergstrom,rudaz,rudsteck}.
In this subsection we combine the latest observational limits on
the photon background with
the calculation by  
Kamionkowski \cite{kamionkowski} and Jungman, Kamionkowski, and
Griest \cite{jungman}
of the flux expected from halo annihilations
to derive a bound in the form of Equation (\ref{n0mubound}).
As discussed in the previous subsection, we only consider annihilations
to two photons of particles with $m_X<0.5~\mev$.

Jungman, Kamionkowski, and Griest \cite{jungman} assume the model
\begin{equation} 
\rho(r)=\rho_0 \frac{R^2+a^2}{r^2+a^2}
\end{equation}
for the dark matter density distribution in the Galaxy, where 
$R$ is the distance of the Sun from the galactic center,
$a$ is the core radius, and $r$ is the distance from the center of
the galaxy.  Note that simulations of cold dark matter models predict
central cusps instead of a core \cite{navarro}, 
which could lead to enhanced annihilation signals toward the center of
the Galaxy; however, we want to be conservative so we do not assume
a cusp.  
Re-writing Equation (10.1) of reference \cite{jungman} to include
the possibility that the particle X does not make up all the dark matter,
we find that the energy flux at $E_o=m_X$ is 
\begin{eqnarray}
\frac{dJ}{dE d\Omega} \simeq f_\gamma
\frac{3.0}{{\rm cm^2~ s ~sr}}
\left(\frac{m_X}{\mev}\right)^{-2} 
\left(\frac{\rho^{0.4}_D}{\Omega_D h^2}\right)^2
I(\psi)
\nonumber \\ \times
\left[\left(\frac{\sigma_0}{6\times 10^{-27} \cm^3 \s^{-1}}\right)
\left(\Omega_{X\bar{X}} h^2 \right)^2\right]~,
\end{eqnarray}
where $I(\psi)\sim 1$ depends on the observation angle $\psi$, 
$\Omega_D \geq \Omega_{X \bar{X}}$ is the contribution to the critical 
density from all kinds of dark matter, and
$\rho^{0.4}_D\sim 1$ is the density of dark matter near the solar radius,
in units of $0.4~ \gev ~{\rm cm}^{-3}$.  We have  
assumed $\rho_X(r)\propto \rho_D(r)$, and used the detector 
energy resolution, $\Delta E/E\sim 0.2$, appropriate for the energy
bins in Fig. 10 of reference \cite{skreekumar}.

The relevant energy range for the halo annihilations we 
are considering, $1-500~\kev$, corresponds to a ``bump'' in
the observed spectrum (see \cite{skreekumar}).
To be conservative, we   
construct a simple bound by comparison with the single power law
$dJ/dE d\Omega \lesssim
0.022 \left(E_o/\mev\right)^{-1.2}{\rm cm^{-2}~ s^{-1} ~sr^{-1}}$,
which is an upper bound on 
the energy flux in the full range $1~\kev \lesssim E_o \lesssim 100~\gev$
\cite{skreekumar}.
Since annihilations to two photons 
produce a line source at $E_o=m_X$, we have the bound
\begin{eqnarray}
& f_\gamma & \left(\frac{m_X}{\mev}\right)^{-0.8}
\left(\Omega_D h^2\right)^{-2} \nonumber \\ & & \times 
\left[\left(\frac{\sigma_0}{6\times 10^{-27} \cm^3 \s^{-1}}\right)
\left(\Omega_{X \bar{X}} h^2 \right)^2\right]< 0.008~,
\end{eqnarray}
which we show in Fig. \ref{config} (the observationally 
favored value for the total mass density is 
$\Omega_D h^2 \simeq 0.15$ \cite{mould,donahue}, but we use 0.3 as 
a more conservative upper bound). 
For particles that survive to the 
present, this bound is generally stronger than the bound from
distortions of the CMB.    
Note that our calculation is a somewhat rough estimate because 
the observational bounds are considerably lower at some energies than the 
power law we adopted \cite{skreekumar},
the energy bin width $\Delta E/E$ is approximate,
and we used $I(\psi)=1$ when 
its value can be higher for observation angles near the galactic center
\cite{jungman}.

For particles with $m_X>0.5~\mev$ we might consider the observation 
of annihilation products other than photons (e.g., positrons);
however, reference \cite{jungman} concludes that 
these signals cannot be used to conclusively rule out dark matter
candidates because of astrophysical uncertainties.

\section{DISCUSSION}

Some of the parameter space
that can be constrained by residual annihilations is covered already
by other kinds of constraints.  To put the annihilation constraints
in perspective we review the primary astrophysical ones.

BBN gives limits on the expansion rate of the universe at 
$T\sim 1~\mev$ (sometimes described as a limit on the effective number
of light neutrinos, e.g., reference \cite{olive}).  
The expansion rate would be
affected by an additional particle with mass in the range where our 
CMB bound is most constraining ($m_X\lesssim \mev$), if the particle's 
number density
at the time of BBN is equal to the thermal equilibrium value.  Therefore, 
the annihilation bound on particles with $m_X\lesssim \mev$ is only 
nonredundant for particles
that were not in thermal equilibrium at the time of nucleosynthesis. 

\begin{figure}
\centerline{\psfig{figure=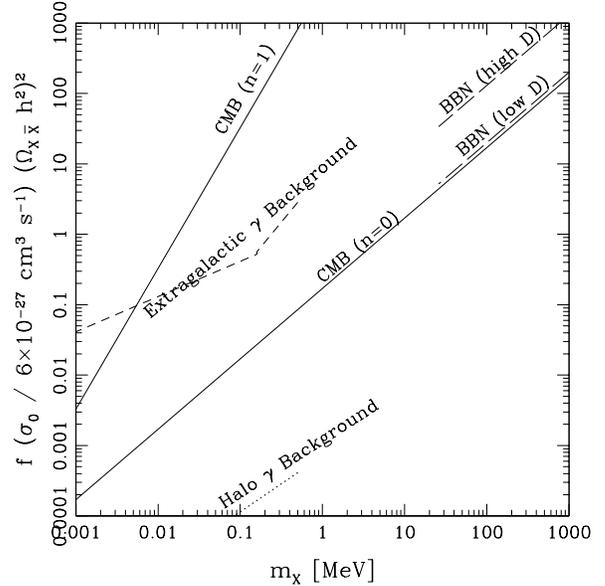,height=8.6cm,width=8.6cm}}
\caption{
Bounds on relic particles from residual annihilations.
The space {\it above} the lines is ruled out in each case.
Except for the upper solid line, all bounds are for particles with 
$\left< \sigma_a \left| v \right| \right> = \sigma_0$ (i.e., $n=0$).
The lower (upper) solid line is the
constraint from the CMB (including both $\mu$ and $y$ distortions)
for $n=0$ ($n=1$).
The upper (lower) long-dashed line is the constraint from BBN using the 
high (low) deuterium abundance (the low D constraint and the CMB
constraint are practically identical but we have introduced a slight offset 
for clarity).
The dashed line is the constraint from the diffuse photon background
produced by extragalactic annihilations, and the dotted line is the 
constraint from the diffuse photon background produced by 
annihilations in the Milky Way halo.   
The BBN constraints apply 
for $m_X>26~\mev$, the threshold for photodissociation of \hef to D.
For the CMB and BBN constraints, $f$ is the fraction of the annihilation 
energy that interacts electromagnetically. 
For the photon background constraints, $f$ should be replaced by
$f_\gamma$, the fraction of annihilations that produce two photons
with energy $m_X$
(the lines for these constraints are terminated at the electron mass
because $f_\gamma<<f$ is likely if other annihilation channels are open). 
For the case of equivalent particle and antiparticle,
substitute $\Omega_{X \bar{X}} \rightarrow \Omega_X$, and 
$\sigma_0/6\times 10^{-27}{\rm cm^3 s^{-1}}\rightarrow  \sigma_0/
3\times 10^{-27}{\rm cm^3 s^{-1}}$. 
}
\label{config}
\end{figure}

Particles with 
$m_X\gtrsim 5~\mev$ are only constrained by our bound  
if their density, extrapolated by $(1+z)^{-3}$ from the
time they influence the CMB to the 
present, is substantially greater than the present critical density, 
or if their cross section is very large but their number density is 
somehow higher than the density obtained from freezeout of their
annihilations (see Fig. \ref{config}).  The first case would 
require that the particle decays
invisibly or otherwise disappears between $z\sim 10^6$ and the present,
and the second requires that the particle was formed by decays of a 
heavier particle (or
some other mechanism) after the freezeout temperature for its annihilations.

New particles that can be created in stars 
(e.g., by plasmon decay, $\gamma \rightarrow X \bar{X}$)
are constrained by their action as additional sources of cooling. 
Constraints from observations of globular cluster stars 
apply for $m_X\lesssim 10~\kev$ \cite{raffelt}.
Similarly, cooling in supernovae, specifically SN1987A, can be influenced 
by the creation of new particles with mass $m_X\lesssim 30~\mev$; 
however, these constraints depend on the details of the particle
interactions in the plasma (e.g., reference \cite{davidson}).

In summary, the bound from distortions of the CMB energy spectrum 
probably cannot be a useful constraint on annihilations of
light ($m_X\lesssim 0.5~\mev$) dark matter particles
(e.g., warm dark matter) to photons, because the 
bound from annihilations in our Galaxy is always stronger.
It is most interesting as a constraint on
{\it dark matter} particles in the mass range 
$0.5\lesssim m_X \lesssim 5~ \mev$, and {\it any} particle that decays
invisibly between $z\simeq 10^6$ and the present, 
although in either case the particle must also evade
the bound from the expansion rate of the universe during BBN.
Finally, although we have not discussed in this paper the case of 
annihilations of particles with cross sections that increase
with decreasing temperature (e.g., \cite{bartlett}), 
it seems likely that they can be very tightly constrained by the 
kinds of tests we have discussed.

\acknowledgements

RJS and TPW were supported by the Department of 
Energy (DE-FG02-91ER40690).

\end{document}